\documentclass[12pt,preprint]{aastex}

\def\aa{{A\&A}}

\def\aj{{AJ}}

\def\apj{{ApJ}}
\def\apjl{{ApJ}}
\def\apjs{{ApJS}}

\def\mnras{{MNRAS}}

\def\fka{{Fe~K$\alpha$ }}
\def\fkb{{Fe~K$\beta$ }}
\def\teng{{Paper~I}}

\received{}
\accepted{}
\slugcomment{Accepted by ApJ: 25 October 2007}
\shorttitle{{\it XMM-Newton} Observation of 1-Jy ULIRG/LINER F04103$-$2838}
\shortauthors{Teng et al.}
\journalid{}{}
\articleid{}{}

\begin{document}

\title{{\it XMM-Newton} Detection of a Compton-thick AGN in the 
1-Jy ULIRG/LINER F04103$-$2838}


\author{Stacy H. Teng{\footnote{Contacting author:
stacyt@astro.umd.edu}}, S. Veilleux, A. S. Wilson} \affil{Department
of Astronomy, University of Maryland, College Park, MD 20742}
\author{A. J. Young} \affil{Center for Space Research, Massachusetts
Institute of Technology, Cambridge, MA 02139}
\author{D. B. Sanders} \affil{Institute for Astronomy, University of
Hawaii, 2680 Woodlawn Drive, Honolulu, HI 96822} \and
\author{N. M. Nagar} \affil{Astronomy Group, Departamento de F\'isica,
Universidad de Concepci\'on, Casilla 160-C, Concepci\'on, Chile}

\begin{abstract}

  We report on the detection of \fka emission in F04103$-$2838,
  an ultraluminous infrared galaxy (ULIRG; log[L$_{\rm IR}$/L$_\odot$]
  $\ge$ 12) that is optically classified as a LINER.  Previous {\it
    Chandra} observations suggested the presence of both a starburst
  and an AGN in this source.  A deeper ($\sim$20~ksec) {\it
    XMM-Newton} spectrum reveals an \fka line at rest frame energy
  $\sim$6.4~keV, consistent with cold neutral iron.  The best-fit
  spectral model indicates the \fka line has an equivalent width of
  $\sim$1.6~keV.  
The hard X-ray emission is dominated by a Compton-thick AGN
  with intrinsic 0.2--10~keV luminosity $\sim$10$^{44}$
  ergs~s$^{-1}$, while the soft X-ray emission is from $\sim$0.1~keV gas
  attributed to the starburst.  The X-ray spectrum of this source
  bears a striking resemblance to that of the archetypal luminous
  infrared galaxy NGC~6240 despite differences in merger state and
  infrared properties.

\end{abstract}


\section{Introduction}
\label{sec:intro}

The primary energy source (AGN versus starburst activity) of
ultraluminous infrared galaxies (ULIRGs; log[L$_{\rm IR}$/L$_\odot
\ge$ 12]) is still a matter of debate.  Optical and infrared
emission-line spectra suggest that the energy output of most local
ULIRGs is dominated by starbursts, but the ``warm'' infrared colors
and quasar-like spectra of the more luminous ULIRGs indicates that
black-hole driven activity plays an increasingly important role in
these objects (e.g., Veilleux et al. 1995, 1997, 1999a, 1999b; Genzel
et al. 1998; Surace \& Sanders 1999; Tran et al. 2001).  The dusty,
gas-rich nature of ULIRGs implies, however, that observations in
energy bands other than radio and X-ray may not always probe the true
nuclear energy source of these objects.  Since the luminosity of
ULIRGs in the radio is insignificant, X-ray observations remain
arguably the best option to solve this energy source mystery.

Unresolved hard X-ray emission is in principle a telltale sign of a
dominant AGN. However, if a large column density of gas ($\ga 10^{24}$
cm$^{-2}$) is located in front of the nucleus, then directly viewed
X-rays from the AGN will be strongly attenuated.  For such cases, \fka
lines with large equivalent widths ($\ga$ 1 keV) are expected due to
scattering off circumnuclear material (e.g., Ghisellini et al.
1994; Krolik 1994). Thus, the discovery of such \fka lines may be the
best evidence for energetically dominant AGNs in highly obscured
ULIRGs.


In recent years, three X-ray surveys have added considerably to our
knowledge of ULIRGs. \citet{ptak} performed a volume-limited
($z < 0.045$) survey of ULIRGs with {\it Chandra}.  On the basis of
their dust temperatures (25-to-60~$\mu$m flux ratio) and X-ray
luminosities, three of the eight ULIRGs sampled by \citet{ptak} were
classified as AGN-dominated (Mrk~231, Mrk~273, and F05189-2524).  In
the same year, \citet{frances} published the results of a similar
survey with {\it XMM-Newton} that focused on the brightest local
ULIRGs (only one ULIRG in their sample had $z > 0.082$).  Of the ten
ULIRGs sampled by \citet{frances}, three were AGN-dominated (Mrk~231,
F19254-7245, and F20551-4250) and two had X-ray signatures of both a
starburst and an AGN (F20100-4156 and F23128-5919).  All of the
AGN-dominated ULIRGs showed strong Fe~K emission lines
(Maloney \& Reynolds 2000; Braito et al. 2003; Ptak et al. 2003; Franceschini et al. 2003; Braito et al. 2004).  These
two pioneering surveys proved the viability of using the X-ray
emission as a diagnostic for AGN activity in ULIRGs.  However, these
two surveys only studied a small set of the nearest and brightest
ULIRGs, and therefore were not able to draw general conclusions on the
issue of the energy source among ULIRGs as a class.

In an attempt to expand this type of study to a more characteristic
sample of ULIRGs, our group (Teng et al. 2005, hereafter \teng)
conducted a snapshot (10 ksec/target) survey of 14 ULIRGs from the
1-Jy sample\footnote{The 1~Jy sample of ULIRGs is comprised of {\it
    IRAS} galaxies with fluxes at 60~$\mu$m exceeding 1~Jy,
  L$\rm{_{IR} \ge 10^{12}}$~L$\rm{_\odot}$, galactic latitude
  $\mid$b$\mid$ $> 30^\circ$, f(60~$\mu$m) $>$ f(12~$\mu$m) (to avoid
  stars), {\it IRAS} color log[$f_{60}/f_{100}$] $>$ --0.3 (to favor
  luminous infrared systems), and
  redshift 0.018 $< \rm{z} <$ 0.268 (Kim \& Sanders, 1998).}.
These sources were carefully selected to sample the full range of
infrared luminosities and infrared colors that characterize the entire
class of local ULIRGs. All 14 galaxies were detected by {\it Chandra},
though most (11/14) had less than 40 counts.  The analysis showed that
the two brightest galaxies in the sample have optical and X-ray
spectral characteristics of Seyfert~1 nuclei.  Most others have X-ray
photon indices (estimated using hardness ratios) and hard X-ray to
far-infrared flux ratios which are similar to those of starbursts.

One exception, F04103$-$2838, had a hardness ratio (deduced from only
30 counts) that suggested the presence of a starburst coexisting with
an AGN.  The low signal-to-noise data could not distinguish between a
Compton-thick AGN or an intrinsically faint nuclear source.  This
object is optically classified as a LINER (Veilleux et al. 1999a).
F04103$-$2838 has one of the largest 25-to-60~$\mu$m flux ratios of
all 1~Jy ULIRGs ($f_{25}/f_{60}$ = 0.30).  In fact, this is the
warmest of all {\it IRAS} 1~Jy ULIRGs with optical LINER or H~II
classification.  This source is even warmer than some of the Seyfert
galaxies in the 1~Jy sample (see Fig. 1 of \teng).  Recent {\it
  Spitzer} detection of [Ne~V] and [O~IV] lines from this source has
confirmed the existence of an AGN in this system (Veilleux et al.
2007, in prep.), making it a rare example of a ULIRG optically
classified as a LINER that is not classified as a starburst on the
basis of mid-infrared spectroscopy (Lutz, Veilleux, \& Genzel 1999;
Taniguchi et al. 1999).  In this paper, we present an {\it XMM-Newton}
observation of F04103$-$2838 which delves deeper into the nature
of this AGN.

F04103$-$2838 is an interacting galaxy system in the late stages of a
merger as indicated by the presence of a single nucleus with distinct
tidal tails \citep{vei02, dasyra}.  This object has an infrared
(8--1000 $\mu$m) luminosity of 10$^{12.15}$L$_\odot$ and a
cosmological redshift of 0.118.  Assuming H$_0$ =
75~km~s$^{-1}$~Mpc$^{-1}$ and q$_0$ = 0 (used throughout this paper),
the luminosity distance of this object is 497~Mpc.  At this distance,
1$\arcsec$ corresponds to $\sim$2.4~kpc. In \S 2 of this paper, we
describe our new {\em XMM-Newton} observation of F04103$-$2838 and the
methods we used to reduce these data. In \S 3, we present the
analysis of these data, emphasizing the results on the X-ray
morphology, the lack of flux variability, and the spectral
decomposition of the X-ray emission. The implications of these results
are discussed in \S 4.  The main conclusions are summarized in \S 5.

\section{Observation and Data Reduction} 
\label{sec:obs}

F04103$-$2838 was observed with {\it XMM-Newton} during orbit \#1132
on 13 February, 2006 (ObsID: 0301330401; PI: Wilson) with the EPIC
instrument.  The EPIC cameras were operating in full-frame mode.  Each
of the detectors used the medium filter.

The data were processed using the standard procedures of the {\it
XMM-Newton} Science Analysis System (SAS) version 6.5.0 released on 17
August, 2005.  The processing procedures outlined in \S 4.11 of the
{\it XMM-Newton} SAS User's Guide were followed.  The event lists were
calibrated with the latest available calibration files as of June,
2006.  Times of high background flares were flagged.  The total good
time interval on source for each camera was 17.5~ksec for PN,
21.8~ksec for MOS1, and 21.5~ksec for MOS2.

Source and background counts were extracted from circular regions with
radii of 24$\arcsec$.  Because the source is near a gap in the CCD and
a nearby X-ray luminous source, the background was extracted from a
circular region with the same area as the source extraction region in
a neighboring piece of the sky in which no obvious X-ray sources reside.
The total 0.2--10~keV counts extracted from the source region are 224
for PN, 52 for MOS1, and 48 from MOS2.  The expected background
  counts in the source region are 45 for PN and 50 for MOS1/2 based on
  the expected background count rates quoted in the {\it XMM} User's Handbook.  Since
  the extracted source counts are approximately the same as the
  expected background counts for the MOS detectors, we will exclude
  the MOS spectra in our spectral analysis of the source.

\section{Analysis}
\label{sec:analysis}

In \S \ref{sec:struc}, we describe the distribution of the X-ray
emission from F04103$-$2838. In \S \ref{sec:var} we point out the lack
of variability of this object. A detailed analysis of the X-ray
spectrum and iron complex is presented in \S \ref{sec:spectra}.

\subsection{Morphology}
\label{sec:struc}

\begin{figure}
\figurenum{1}
\epsscale{1.0}
\plotone{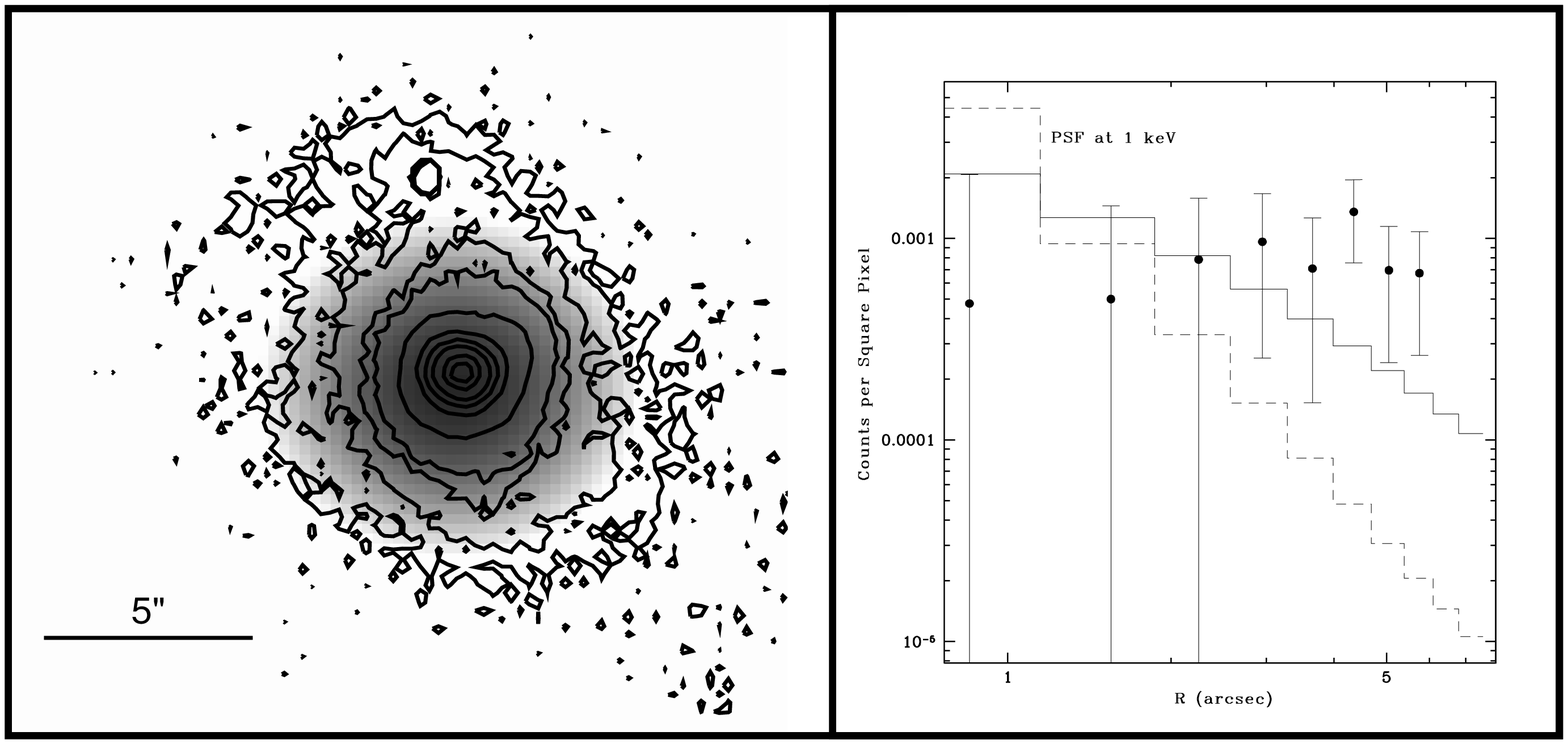}
\caption{{\it Left}: EPIC mosaic image of F04103--2838 smoothed with a
  Gaussian (FWHM$\sim$5$\arcsec$) and displayed on a linear grey scale.  The
  contours are optical R-band data from \citet{kim02}.  {\it Right}:
  Comparison of unsmoothed EPIC PN PSF at 1~keV with observed 0.2-2~keV radial
  profile of F04103--2838.  The dashed line represents the theoretical PSF
  while the solid line represents the PSF broadened due to
  uncertainties in the correction for pointing drift of the telescope.  The absolute pointing drift (APD) error is
  conservatively assumed to be 3$\arcsec$,
  the upper limit (see {\it XMM-Newton} Observer's Handbook).  The error bars were calculated using
  \citet{stat}.  The source is unresolved within the uncertainty of
  the measurements.}
\label{fig:optcon}
\end{figure}

To improve the signal-to-noise ratio of the images, the PN and MOS1/2
events were combined using the SAS task {\it emosaic} and then
smoothed with a 5$\arcsec$ Gaussian using {\it asmooth} to match the
spatial resolution of {\it XMM}.  The resultant image is
displayed in the left panel of Figure~\ref{fig:optcon}.
A comparison of the 0.2--2~keV (unsmoothed) radial profile with the {\it
  XMM-Newton} point spread function (PSF) at 1~keV indicates that the
source is unresolved (see Fig. 1, right
panel).  Only
the EPIC PN data were used for the radial profile calculations because
of the small number of counts detected by the MOS1/2 cameras.  The PSF
of the PN camera is well described by a King profile\footnote{PSF =
  A$[1 + (\frac{r}{r_0})^2]^{- \alpha}$, where A $\sim$4.756, $r_0 \sim$ 5.5 pixels, and $\alpha
  \sim$ 1.6 \citep{cal}.} and was normalized so that the total number
of counts per square pixel under the curve match the total number of
detected counts per square pixel. The {\it Chandra} data from
  \teng~verifies that the source is unresolved.

\subsection{X-Ray Variability}
\label{sec:var}

\begin{figure}
\figurenum{2}
\epsscale{0.7}
\plotone{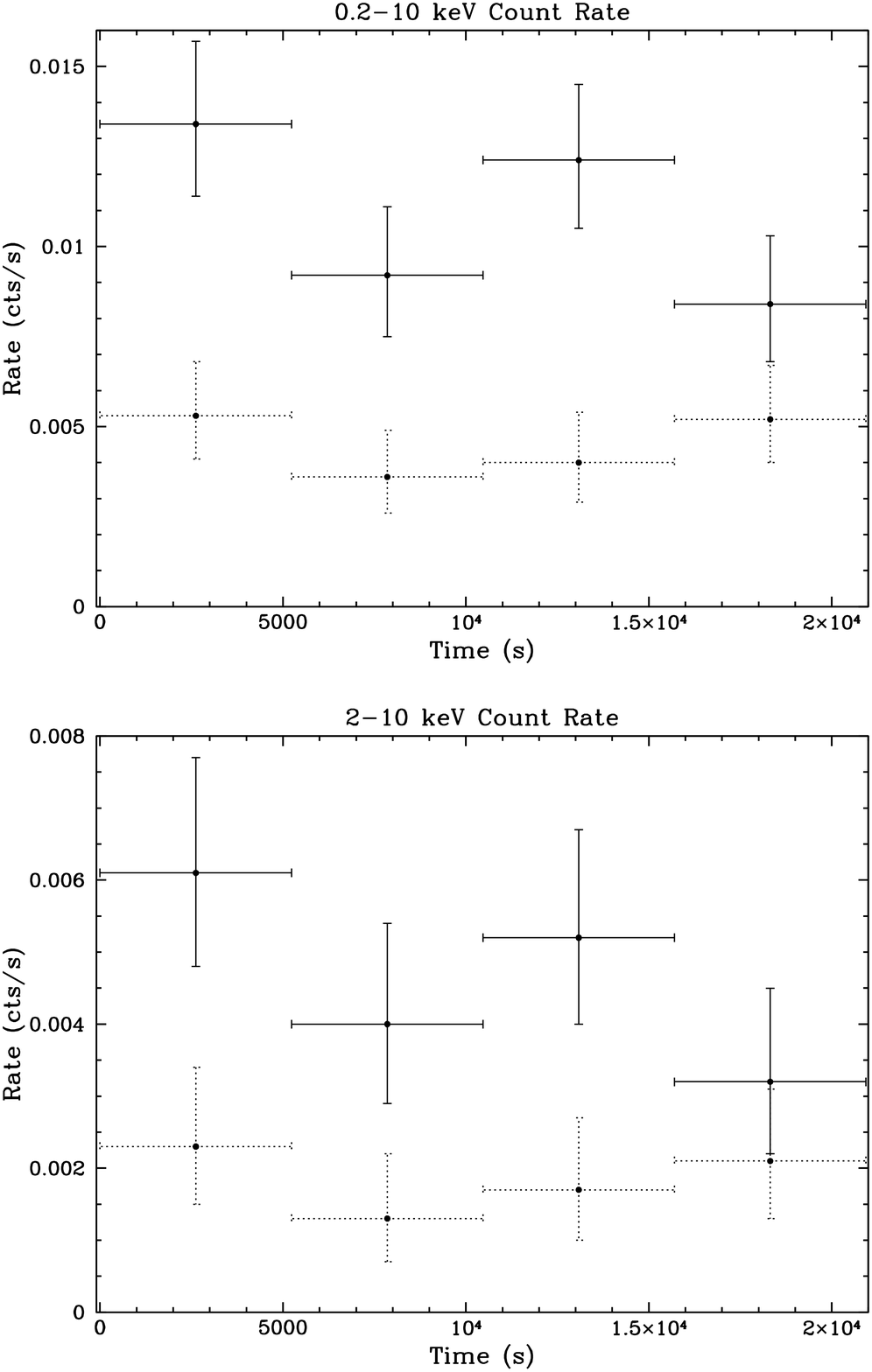}
\caption{The 0.2--10~keV and 2--10~keV light curves for F04103$-$2838.
  The solid crosses denote the source count rate, while the dotted
  crosses denote the background count rate.  Background-subtraction
  was not applied to the source spectrum.  The time bins are each 5234
  seconds.  The error bars are Poissonian counting errors calculated
  following \citet{stat} at the 84\% significance level.  Within the
  errors, the source is not variable on the $\sim$20~ksec time scale
  of our observation.  The lack of short timescale variations is
  expected of sources where most of the primary X-ray flux is absorbed
or reprocessed.}
\label{fig:pnlc}
\end{figure}

The time interval covered by our observation was divided into four
equal bins of 5234 seconds to search for significant X-ray variability,
another potential indicator of dominant AGN activity.  The 0.2--10~keV
and 2--10~keV EPIC PN count rates were calculated for both source and
background.  Figure~\ref{fig:pnlc} shows the 0.2--10 and 2--10~keV
light curves of the source and background.  To within the errors, the
source is not significantly variable on the 5-6 hour time scale of our
observations.

\subsection{X-Ray Spectra}
\label{sec:spectra}

The extracted source and background spectra from each detector were binned using the
FTOOL {\it grppha} to at least 3, 5, and 15
counts bin$^{-1}$.   The binned and unbinned spectra were then analyzed using XSPEC version 11.3.2t.  The quoted errors on
the derived best-fitting model parameters correspond to a 90\%
confidence level ($\Delta \chi^2/\Delta$c-stat = 2.706).   The $\chi^2$ goodness-of-fit test was used to
judge the fits to the spectrum binned to at least 15 counts
bin$^{-1}$.  The Cash statistics (c-stat) option in XSPEC was used for
spectra binned to at least 3 and 5 counts bin$^{-1}$ and the unbinned
data.  The spectral model was applied to the EPIC PN data only
  (see \S \ref{sec:obs}).  All models were corrected for Galactic absorption using
N$_{\rm H, Galactic} = 2.45 \times 10^{20}$~atoms~cm$^{-2}$ \citep{nh}.

\subsubsection{Effects of Binning}
\label{sec:bins}

By definition, spectra binned to at least 15 counts bin$^{-1}$ have the highest
signal-to-noise ratios while the spectra binned to at least 3 counts
bin$^{-1}$ show the most spectral details.  The first task is to
determine whether the mode of binning affects the spectral parameters
derived from the best-fit model\footnote {Gaussian statistics apply
  to data binned to at least 15 counts bin$^{-1}$ while Poisson
  statistics apply to the data binned to at least 3 or 5 counts
  bin$^{-1}$ and unbinned data.  Since the difference of two Gaussian distributions remains a Gaussian distribution, a background-subtracted
spectrum binned to at least 15 counts bin$^{-1}$ retains the
properties of a Gaussian distribution and can be modeled normally.
However, the same is not true for a Poisson distribution.  Therefore,
the background cannot be simply subtracted for data binned to at least
3 or 5 counts bin$^{-1}$ and unbinned data and then modeled.  One way of treating the
background is to model the background spectrum separately and then add
the background model to the continuum model when fitting the source
spectrum.  For this paper, the background is modeled using a simple,
relatively flat power law ($\Gamma \sim$1.0).  This treatment of the background is
applied to all modeling of data binned to at least 3 and 5 counts
bin$^{-1}$ and the unbinned spectrum.  A representation of the
background spectrum and model is shown in the bottom panel of
Figure~\ref{fig:bestfit}.}.  Since Cash-statistics were developed
for the modeling of unbinned data, we also modeled the unbinned
spectrum for comparison.

\begin{figure}
\figurenum{3}
\epsscale{1.0}
\plotone{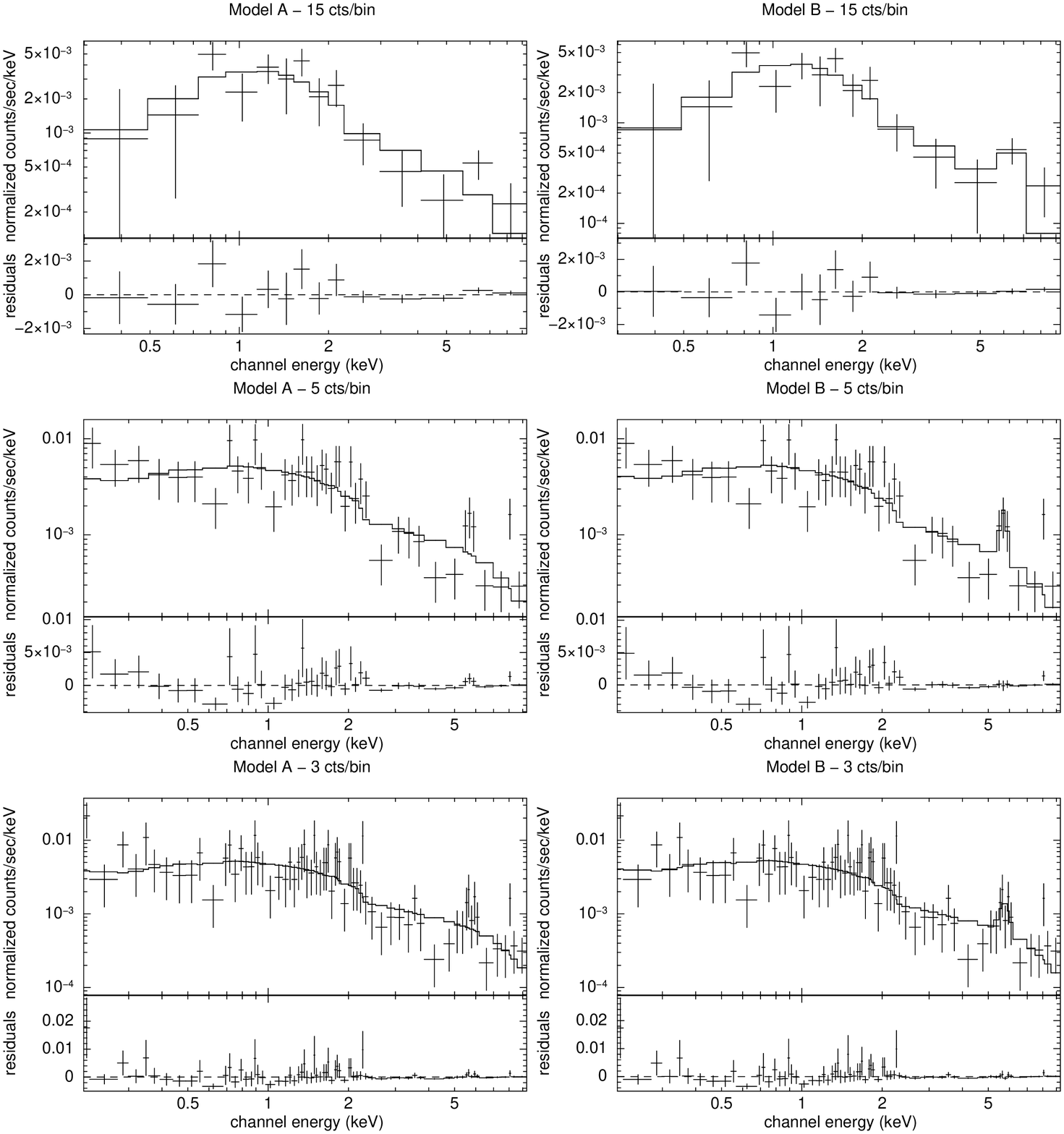}
\caption{EPIC PN spectrum and best-fit models to F04103$-$2838 with different
  binnings: $\ge$ 15 counts bin$^{-1}$ (top panels), $\ge$ 5 counts
  bin$^{-1}$ (middle), and $\ge$ 3 counts bin$^{-1}$ (bottom).  The
  unbinned spectrum was modeled but is not shown here.  Model A (left
  panels) is a simple absorbed power-law distribution; model B (right)
  is the same as A, but includes a Gaussian component to model the
  Fe~K emission.  The best-fit model parameters are listed in
  Table~\ref{tab:binfits}.  The iron line is most prominent in the data binned to
  5 counts bin$^{-1}$ and the Fe~K doublet may be present in the
  3 counts bin$^{-1}$ data.  X-axis of the figures represents energy in the
  observer's frame.}
\label{fig:binspecs}
\end{figure}

\begin{figure}
\figurenum{4}
\epsscale{0.7}
\centering
\plotone{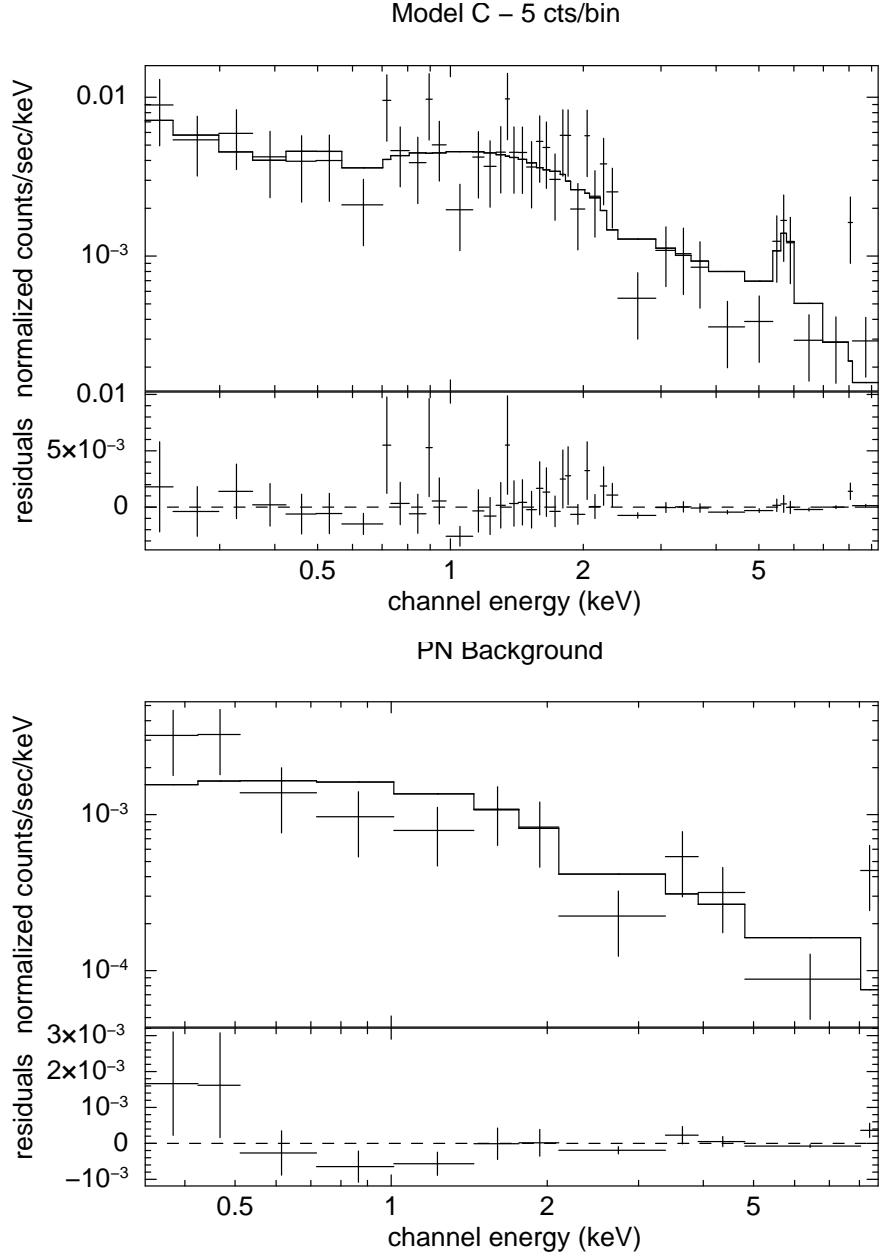}
\caption{EPIC PN source and background spectra of F04103$-$2838
    binned to at least 5 counts bin$^{-1}$ with the best-fit {\it
      unbinned} model (C) applied.  The X-axis of the figures
    represents energy in the observer's frame.  {\it
    Top:} The source model includes a MEKAL component for the thermal
  emission, a power-law component to represent the AGN component, a Gaussian
  component to model the Fe~K emission, and a relatively flat
  power-law ($\Gamma \sim$1.0) for the background.  The model
  parameters are listed in Table~\ref{tab:binfits}.  {\it Bottom:} A binned
  background spectrum with the background model used in the modeling
  of the unbinned spectrum and spectra binned to at least 5 and 3 counts
  bin$^{-1}$.  No significant features are seen in the background spectrum.}
\label{fig:bestfit}
\end{figure}

Two simple models were applied to the spectra.  Model A is an absorbed
power-law distribution.  Model B is the same as A, except for the
inclusion of a Gaussian component to model the Fe~K emission at
6--7~keV (rest frame).  Table~\ref{tab:binfits} lists the best-fit
parameters of each model and Figure~\ref{fig:binspecs} shows each set
of spectra with the best-fit models.  The significant improvement in
fitting statistics of model B over model A suggests that there is
indeed an emission line at an energy consistent with \fka emission.
However, since the number of counts is relatively low (especially when
the data is binned to only 3 or 5 counts bin$^{-1}$), the F-test
cannot be used to determine whether the addition of the Gaussian
component to model A is significant.  The likelihood of the line being
a result of statistical variations was tested using simulations.
To this end, 10000 spectra were created using the {\it fakeit} command
in XSPEC for each set of binned or unbinned data.  The simulated spectra were
created using model A.  Then these spectra were fitted by both models
A and B.  If the line is a result of statistical variations, then one
would expect a large fraction of the simulated spectra to be well described
by model B.  The fitting statistics were used to calculate
$\Delta$c-stat(A--B) [or $\Delta \chi ^2$(A--B) for the 15 counts
bin$^{-1}$ data] which was then compared with the values presented in
Table~\ref{tab:binfits}.  For the 15 counts bin$^{-1}$ data, 1000 of
10000 (10.0\%) had $\Delta \chi^2$ greater than 3.76.  This implies that
model B (the inclusion of the emission line) is significant at the
90.0\% level (a 1.6-$\sigma$ detection).  Similarly, the simulations show that the line is
significant at the 96.87\% level (313 out of 10000; 2.2$\sigma$) for the 5 counts bin$^{-1}$ data, at the 93.5\% level
(507 out of 10000; 1.8$\sigma$) for the 3 counts bin$^{-1}$ data,
  and
at the 94.0\% level (608 out of 10000; 1.9$\sigma$) for the unbinned
data.  From these simulations, the line is significant to at least the
90.0\% level.

The 3 counts bin$^{-1}$
data also suggest that the iron line can be decomposed into two
narrower emission lines with centroid energies at 6.3 (EW
$\sim$0.6~keV) and 6.7~keV (EW $\sim$0.4~keV) in
the rest frame.  These energies are consistent with emission arising
from neutral iron and Fe~XXV, respectively.  The fitting statistics of
  the double-line model to the unbinned data is only slightly better
  than that of the single-line model.  The detection of these narrow
  lines in the Fe~K complex is significant at only the $\sim$60\%
  level based on 10000 simulations of the unbinned data.  Therefore,
  the detection of the doublet needs to be confirmed with data of
  higher spectral resolution and signal-to-noise ratio.

Our modeling and simulations show that Cash-statistics give
  consistent results for the unbinned spectrum and the spectra binned
  to at least 3 and 5 counts bin$^{-1}$.  Since Cash statistics were
  designed for unbinned spectra, we will use only the unbinned
  spectrum in subsequent modeling.  The iron line is most prominent in
the data binned to at least 5 counts bin$^{-1}$, we will use the
spectrum binned to at least 5 counts bin$^{-1}$ as a visual and
qualitative check for the model of the unbinned data.

\subsubsection{AGN + Starburst Continuum Models}
\label{sec:model}

Aside from models A and B mentioned above, we modeled the unbinned spectrum with slightly more complex models to account for the
possibility that a starburst may coexist with the AGN in
F04103$-$2838.  Cautioning against over-interpreting data with only
modest signal-to-noise ratios, even these more ``complex'' models were
kept as simple as possible.

The first model (model C) is a combination of absorbed power-law and
MEKAL spectra (with metallicity fixed at solar) representing the
emission from the AGN and starburst, respectively.  The second model
(model D) is a combination of two absorbed power laws, with one power
law representing the AGN and the other representing the high mass
X-ray binaries (HMXBs) associated with the possible starburst in this
object. Finally, a third model (model E) was a combination of the two
above mentioned models: a power law for the AGN, a power law for the
HMXBs, and a MEKAL model for the hot gas.  For all of these models, a
Gaussian with centroid energy between 6 and 7~keV was included to
model the iron line.  

While all of these models give better fitting statistics
than the simpler power law models, only model C is a realistic fit to the
data.  Models D and E are rejected on the grounds that the best-fit
power law values are physically unrealistic descriptions of AGNs.  Therefore, we adopt model
C as the ``best-fit model'' (Figure~\ref{fig:bestfit}) and list the
fitting parameters in Table~\ref{tab:binfits}.  This is perhaps
not surprising given that ULIRGs are known from observations at
optical and infrared wavelengths to show the presence of both an AGN
and a starburst (e.g., Genzel et al. 1998; Kim, Veilleux, \& Sanders
1998); F04103$-$2838 does not appear to be an exception.

\section{Discussion}
\label{sec:dis}

\subsection{The Soft Component}
\label{sec:soft}


The results from the spectral fitting suggest that the soft X-ray
(0.2--2~keV) flux is best described as thermal emission from hot gas
with kT $\sim$ 0.1~keV (T$\sim$1.2$\times 10^6$~K).  This is somewhat
lower than the range of gas temperatures (0.6--0.8~keV) found in LINERs
\citep{gonzalez}.  The results for F04103--2838
is also somewhat lower than the results from \citet{grimes} who
performed a {\it Chandra} archival study of the soft X-ray emission
from starburst galaxies ranging in luminosity from dwarf
galaxies to ULIRGs.  The authors found that the soft X-ray thermal
emission of these starburst galaxies tends to fall in the temperature
range kT $\sim$0.25--0.8~keV with ULIRGs occupying the upper end of
this temperature range.  
These large temperatures can all be attributed to powerful starbursts.

The soft X-ray emission in F04103$-$2838 is likely the result of
thermal bremsstrahlung from a hot gas produced by the merger-induced
starburst or by intrinsically extended soft X-ray emission heated by
the AGN.  
If the ion density equals that of the electrons, the relationship between the
electron density ($n_e$) and luminosity of an emitting region of a
given volume ($V$) is
\begin{equation}
L_{ff} \approx 1.7 \times 10^{-25} n_e^2 f V~{\rm ergs~s^{-1}},
\label{eq:brems}
\end{equation}
where $f$ is the filling factor for the hot
gas\footnote{Equation~\ref{eq:brems} is based on equation (5.14b) and
  Figure~5.2 of \citet{rybicki} for T = $10^6$~K in the energy range
  of 0.2--2~keV.}.  The non-AGN
contribution of the nominal 0.2--2~keV luminosity from the best-fit
model (model C) for F04103$-$2838 is $1.6 \times 10^{41}$~ergs~s$^{-1}$.
Assuming the emitting region is spherical with a diameter of
$\leq$5$\arcsec$\footnote{While the selection of a $\leq$5$\arcsec$ emitting
  region is based on the spatial resolution of the telescope, it
  should be noted that the linear diameter of 5$\arcsec$ at the
  distance of F04103--2838 is less than a factor of two larger than
  the soft X-ray (0.5--2.5~keV) emitting region of NGC~6240
  \citep{komossa}.  Therefore, the assumption of a $\leq$5$\arcsec$
  diameter is reasonable, even though it was chosen based on the
  instrument PSF.}, the
average electron density has a lower limit of
$\sim$0.19$f^{-1/2}$~cm$^{-3}$.  This value is consistent with
simulation results for the warm (10$^{5.5} \lesssim$ T
$\lesssim 10^{6.5}$~K) component in the wind models of \citet{winds}.

Observationally, this hot gas component is difficult to probe because
of its low density and emissivity.  \citet{winds} performed
hydrodynamic simulations of starburst-driven galactic winds with
various ISM models.  The authors found that, in general, the soft
X-ray emission comes from gas with low filling factors ($10^{-3} < f
< 10^{-1}$; see also Cecil, Bland-Hawthorn, \& Veilleux 2002; and Strickland
et al. 2004a, 2004b for observational constraints).  Using these values for $f$, the electron
density of the hot gas in F04103$-$2838 is $\sim$0.6--5.9~cm$^{-3}$,
consistent with values derived by \citet{netzer} in NGC~6240.

The soft X-ray emission detected in F04103--2838 may be thus the
result of superwinds from the starburst.  X-ray superbubbles have been
observed in Arp~220 \citep{arp220} and NGC~6240 \citep{netzer}.
Furthermore, powerful outflow events are now thought to take place in
most ULIRGs (e.g., Rupke et al. 2002, 2005a, b, c, though their sample did not
include F04103--2838).

\subsection{The Iron Feature}
\label{sec:linedis}

F04103$-$2838 joins the growing list of ULIRGs with Fe~K detections
[e.g. Arp~220 \citep{arp220}, Z11598-0112 (\teng), F19254-7245
\citep{frances, braito03}, Mrk~231 \citep{malreynolds, ptak,
  braito04}, F05189-2524 \citep{ptak}, Mrk~273 \citep{ptak}, and
UGC~05101 \citep{iman, ptak}], supporting the view that an obscured
AGN exists in many of these objects. The presence of an AGN in
F04103$-$2838 was first suggested by \teng\ based on the large hard
X-ray to far-infrared flux ratio; the {\em XMM} detection of Fe~K now
indicates that the luminosity of this AGN has probably been
underestimated.

Few LINERs have detected \fka lines.  \citet{tera} studied a sample
of 53 LINERs and low-luminosity Seyfert galaxies using {\it ASCA}.  Of
the 21 LINERs in their sample, Fe emission lines were detected in only five
galaxies (NGC~1052, NGC~3998, NGC~4261, NGC~4579, and NGC~4736). Of
these five objects, only four (i.e. those excluding NGC~4261) have centroid line
energies consistent within the uncertainties of the measurements with
\fka emission due to neutral iron (E $\sim$ 6.4~keV).

Three other LINERs have known Fe~K detections; all three are powerful
luminous or ultraluminous infrared galaxies.  These galaxies are
Arp~220 \citep{arp220}, NGC~6240 \citep{ptak, komossa}, and UGC~5101
\citep{iman, ptak}.  {\it Chandra} observations of Arp~220, the
archetypal ULIRG, show an iron line at 6.7$\pm$0.1~keV.
This is consistent with emission due to Fe~XX up to Fe~XXVI, but not
neutral iron at 6.4~keV \citep{arp220}.  \citet{komossa} detected Fe~K emission from each of the two nuclei in NGC~6240.  Their analysis showed that the iron lines in each nucleus are consistent with Fe K$\alpha$ and Fe K$\beta$ emissions.

\begin{figure}
\figurenum{5}
\epsscale{0.75}
\centering
\rotatebox{-90}{\plotone{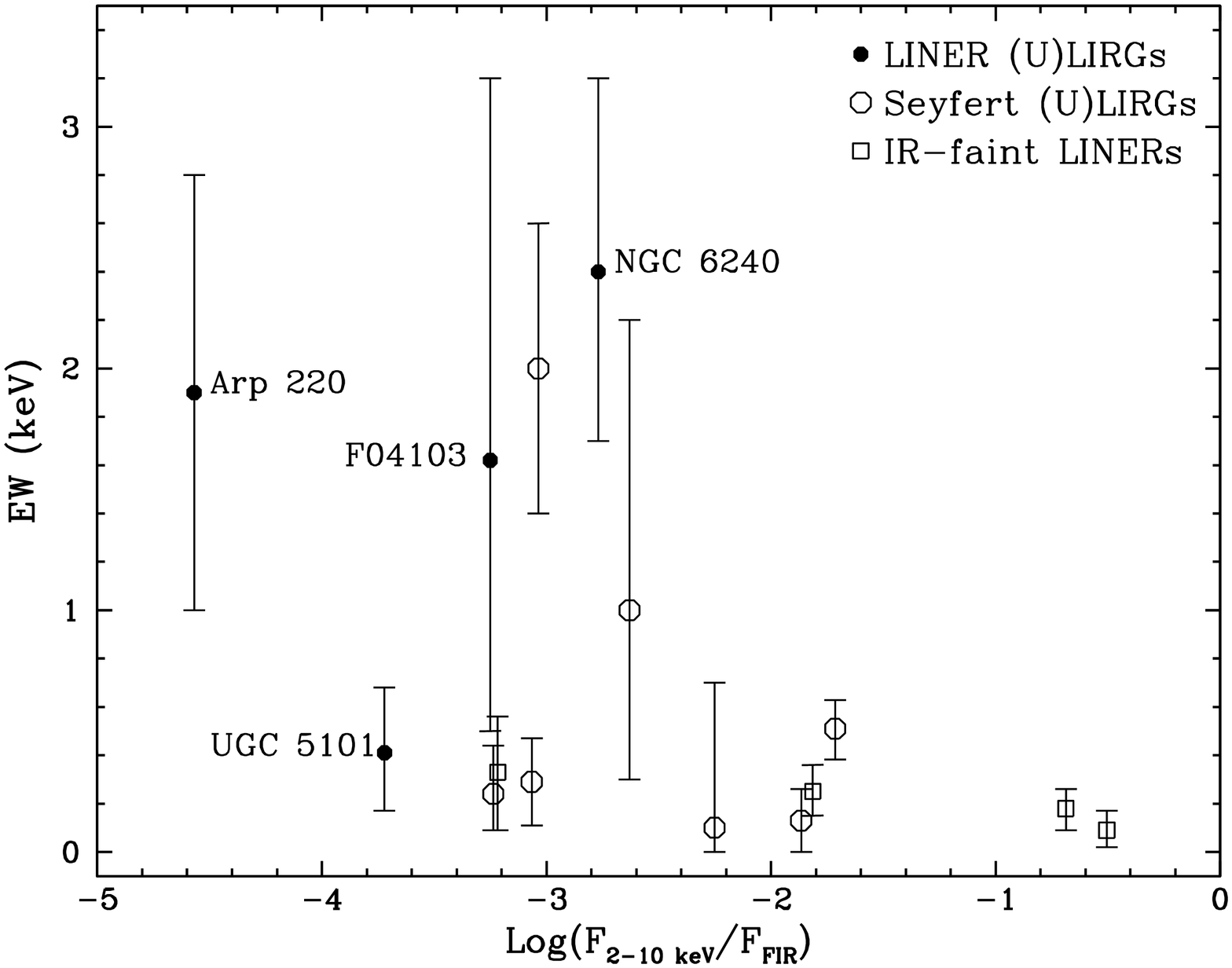}}
\caption{Distribution of equivalent widths of Fe~K emission features in
  LINERs and (U)LIRGs versus absorption-corrected 2--10~keV to far-infrared flux ratio.  In
  general, the (U)LIRGs have higher EWs than IR-faint LINERs.  Of these, F19254--7245 (EW$\sim$2~keV) was
  determined to be Compton-thick by \citet{braito03}.  The values
  included in this figure are drawn from
  \citet{tera}, \citet{braito03}, \citet{iman}, \citet{ptak}, \citet{braito04}, \citet{iman04},
  \citet{arp220}, \teng, and this work.  The equivalent widths for the four
  LINER (U)LIRGs included in this sample are labeled.  The
  F04103--2838 value plotted is derived from our best-fit model (C).  Note that
  \citet{komossa} detected Fe~K emission from each of the two nuclei
  in NGC~6240.  Therefore, the single value quoted by \citet{ptak} is
  a sum of the \fka and \fkb emission due to neutral iron, likely dominated by the brighter
  southern nucleus.  The value quoted for Arp~220 may also be due to a
  blend of emission lines arising from ionized iron (Fe~XX up to
  Fe~XXVI).  The value ($\sim$0.5~keV) quoted for the ULIRG Mrk~463 is
  a sum of the
  emission due to neutral iron and Fe~XXV.}
\label{fig:eqwid}
\end{figure}

In Figure~\ref{fig:eqwid} we show the distribution of published Fe~K equivalent
widths of all LINERs and ULIRGs known to
have line emission.  Arp~220, NGC~6240, and F04103--2838 appear to
have iron emission with the greatest EW measurements of all the LINERs
and ULIRGs.  These large Fe~K features could be
results of the blending of multiple narrower lines.  \citet{komossa} did not
publish the EWs of the lines from each of the nuclei in NGC~6240.  The
result quoted here is from \citet{ptak}.   The authors did not distinguish \fka emission from \fkb
emission and the EW measurement is likely dominated by the brighter southern
nucleus alone.  The large
equivalent widths of the ULIRGs are telltale signs of obscured AGNs where
line-of-sight columns of material exceeding $10^{24}$ cm$^{-2}$
prevent a direct view of the AGN; the 2 -- 10 keV flux is dominated by
light scattered off dust or electrons (e.g., Ghisellini et al. 1994;
Krolik et al. 1994).  The large amount of molecular gas ($\sim$ 10$^4$
M$_\odot$ pc$^{-2}$) within 400 pc from the nuclei of NGC~6240 (e.g.,
Bryant \& Scoville 1999) is sufficient to cause this obscuration. A
similar explanation likely applies to F04103$-$2838, although we are
not aware of any CO measurements in this system.  


Interestingly, the Fe~K complex in NGC~6240 breaks up into a number of
narrow lines.  Both \citet{netzer} and \citet{boller03} detected Fe~K
lines due to neutral iron (6.41$\pm$0.2~keV), Fe~XXV
(6.68$\pm$0.02~keV), and Fe~XXVI (7.01$\pm$0.04~keV) in NGC~6240.  \citet{komossa} also detected lines at 6.4 and 6.95~keV.  The
centroid energies of the lines due to neutral iron and Fe~XXV in
NGC~6240 are consistent with the respective centroid energies suggested by the
doublet in the F04103$-$2838 3 counts bin$^{-1}$ data.  Although
simulations suggest the two line model is only significant at the
$\sim$60\% level, a FWHM of
$\sim$30000~km~s$^{-1}$ ($\sigma \sim$0.3~keV) seems too broad and the
two component interpretation may be more likely.  The Fe~XXVI line in NGC~6240 is much fainter than
the other lines so it is not surprising that we were unable to
detect this feature in the modest signal-to-noise ratio data of
F04103$-$2838.

Despite their X-ray similarities, F04103$-$2838 is $\sim$2.5 times
more infrared luminous than NGC~6240.  These objects also differ in
terms of {\it IRAS} $f_{25}/f_{60}$ ratios (0.15 for NGC~6240 and 0.30
for F04103$-$2838) and merger state (NGC~6240 is in a pre-merger phase
with a nuclear separation of $\sim$ 1.3 kpc while F04103$-$2838 is in
the post-merger stage with a single coalesced nucleus). There is growing observational evidence (e.g., Veilleux et
al. 2002, 2006; Ishida 2004; Dasyra et al. 2006a, b, 2007) and theoretical motivation (e.g., Hopkins
et al. 2005) that mergers of gas-rich galaxies often
produce ``cool'' ($f_{25}/f_{60} < 0.2$) luminous infrared galaxies
that evolve into ``warm'' ($f_{25}/f_{60} \ge 0.2$) ULIRGs before
becoming optical quasars.  If this evolutionary sequence applies to
NGC~6240 and F04103$-$2838, the first object may actually be the
precursor to the latter.

\subsection{Energy Source of the ULIRG}
\label{sec:energy}

The lack of of short timescale variability (see \S \ref{sec:var}) is to be expected if most
of the primary X-ray flux is being absorbed or reprocessed.  As discussed in \S 4.2, the large equivalent width of the iron line in
F04103$-$2838 implies the presence of a highly obscured AGN.  It is
very difficult in such cases to estimate the intrinsic luminosity of
the AGN without measurements of the $>$ 10 keV flux from the buried
AGN (e.g., Mrk~231; Braito et al. 2004).  Here we follow the method of
\citet{malreynolds} to estimate the intrinsic luminosity of F04103--2838.

In their analysis of an {\it ASCA} observation of Mrk~231, they
discussed two ways of estimating the intrinsic AGN flux.  The observed
X-ray flux is due to a combination of two effects: reflection and
scattering.  \citet{malreynolds} estimated the intrinsic AGN flux from
the reflection and the scattering components separately.  In their
geometry, the observer has an obstructed view of the
nucleus so the observed flux must be either scattered or reflected
into the line of sight along which there is some amount of absorbing material.  The reflected component is light from the central engine
reflected off of the circumnuclear torus; the amount of reflection
depends on the size of the reflecting surface and the composition of
the torus.  On
the other hand, the scattered component is light from the central
engine (unobstructed by the torus) scattered into the line of sight.
Based on their spectral fitting of the {\it ASCA} data,
\citet{malreynolds} found that the X-ray flux of Mrk~231 is scattering-dominated with 75\% scattered and 25\% reflected light.

Due to the low signal-to-noise ratio of our data on F04103--2838, the
same spectral fitting as performed by \citet{malreynolds} could not be
done.  The large equivalent width of the Fe~K$\alpha$ line
($\sim$1.6~keV) above 1~keV suggests a reflection-dominated spectrum.
However, the width of the line implies it could be a blend of narrower
Fe~K$\alpha$ and ionized iron emission lines (as suggested by the
3~counts bin$^{-1}$ data).  If this were the case, the \fka EW may be
more consistent with a scattering dominated spectrum.
Therefore, we
will consider two cases: (1) the majority of the observed flux is due
to reflection and (2) the majority of the observed flux is due to
scattering
to estimate the intrinsic X-ray luminosity of the AGN.  

After correction for absorption, the nominal 0.2--10~keV flux of the
buried AGN in F04103--2838 derived from our best-fit model (model C)
is $1.83 \times 10^{42}$ ergs~s$^{-1}$.  In the first
scenario, we will assume the reflection component is 75\% and the
scattering component is 25\% of the total observed flux.  This implies
that  $L_{\rm scattered} = 0.45 \times 10^{42}$ ergs~s$^{-1}$ and
$L_{\rm reflected} = 1.38 \times 10^{42}$ ergs~s$^{-1}$ for the AGN in
F04103$-$2838.  In \citet{malreynolds}, the luminosity from the
reflected portion is scaled up by a factor of 25 in their modeling of
the reflection process.  The reflection process differs for different
galaxies; it depends on the ionization state of the mirror and the
steepness of the photon index of the central black hole.
\citet{malreynolds} assumed reflection from neutral material, a
reflecting fraction of 10\%, and the
canonical value of the photon index due to an AGN ($\Gamma$ = 1.8).
The scaling factor used by \citet{malreynolds} corrects for the
flattening of a spectrum with $\Gamma$ = 1.8 to $\Gamma \sim$1.1 (based on a single
absorbed power-law model) for Mrk~231 due to reflection.  
The correction factor of 25, therefore, is a maximum correction factor.
The minimum scaling factor is 10 to simply correct for a reflecting surface
fraction of 10\%.  
We will conservatively assume this minimum scaling
factor of 10 for the reflection component.
For the scattering
component, we will assume the same scattering fraction as
\citet{malreynolds} (i.e. 1\% for electron scattering).  After the corrections,
the intrinsic 0.2--10~keV luminosity of the AGN in this scenario, where
reflection dominates the observed flux, is $1.4 \times 10^{43}$ ergs~s$^{-1}$ (from
reflection) and $4.5 \times 10^{43}$ ergs~s$^{-1}$ (from scattering).
Therefore, the total reflection- and scattering-corrected luminosity in the
0.2--10~keV band is $5.9 \times 10^{43}$~ergs~s$^{-1}$ if we assume
reflected light dominates the observed spectrum. 

Similarly, for the second scenario where the majority of the observed
flux was scattered into the line of sight, we will
assume the reflection component is 25\% and the scattering component
is 75\% of the total observed flux.  This implies that the intrinsic
0.2--10~keV luminosity of the AGN is $4.5 \times 10^{42}$ ergs~s$^{-1}$ (from
reflection) and $1.4 \times 10^{44}$ ergs~s$^{-1}$ (from scattering).
Hence, the corrected 0.2--10~keV luminosity is $1.4 \times
10^{44}$~ergs~s$^{-1}$.  It should be noted that in both cases we
considered, the luminosity from the scattering portion dominated the
total after the corrections.


Thus, with the assumptions made above, the intrinsic 0.2--10~keV
luminosity of the AGN ranges from 0.6 to $1.4 \times 10^{44}$
ergs~s$^{-1}$.  This range in luminosity overlaps with that of quasars ($\sim$10$^{44}$~ergs~s$^{-1}$; e.g.,
Elvis et al. 1994; Piconcelli et al. 2005) and is similar to that of NGC~6240 ($\sim$ 0.7 $-$ 2 $\times$ 10$^{44}$
ergs s$^{-1}$, after correction for an HI column density of 1 $-$ 2
$\times$ 10$^{24}$ cm$^{-2}$; Vignati et al. 1999).  
The ratio log[L$_{\rm 2-10~keV}$/L$_{\rm IR}$] for F04103--2838
corrected for scattering and reflection is
--2.2 to --1.7.  These values fall precisely within the range found in
radio-quiet PG quasars (--3 to --1; Sanders et al. 1989).  
Assuming F04103--2838 has the same X-ray to bolometric luminosity
ratio as radio-quiet QSOs \citep[L$_{\rm x}$/L$_{\rm bol} \sim$3\%]{elvis}, the AGN contribution to
the bolometric luminosity of F04103--2838 is $\sim$15--38\%.  Therefore,
within the large uncertainties, the AGN in F04103--2838
does not dominate the total energy output of the galaxy.

\section{Summary}
\label{sec:summary}

The results from our analysis of the {\it XMM-Newton}\ spectrum of the
1-Jy ULIRG/LINER F04103$-$2838 can be summarized as follows:

\begin{enumerate}

\item The soft (0.2--2~keV) X-ray flux of F04103$-$2838 is attributed to hot
  gas with kT $\sim$0.1~keV.  This temperature is similar to that
  derived in other starburst galaxies and LINERs.  
  The electron density in F04103--2838 
is $\sim$0.6--5.9~cm$^{-3}$, consistent with theoretical predictions and
  observational estimates in wind systems.

\item An \fka line located at $\sim$ 6.4 keV with an equivalent width
  of $\sim$1.6~keV is detected in F04103$-$2838.  The line could be
  intrinsically broad or could be made up of two narrow lines located at rest frame
  energies of $\sim$ 6.3 and 6.7~keV 
   but this decomposition is only significant at the
  $\sim$60\% level, so it needs to be verified with higher resolution spectra.

\item The large equivalent width of the \fka line suggests that the
  AGN is Compton-thick. Using simple assumptions, we estimate that the
  intrinsic 0.2--10~keV luminosity of this AGN is 0.6 -- 1.4 $\times
  10^{44}$ ergs~s$^{-1}$.  If these assumptions are correct and the
  galaxy has a QSO-like X-ray to bolometric luminosity ratio, the AGN
  detected by our observations does not dominate the bolometric luminosity of
  F04103$-$2838.

\item The X-ray spectral characteristics of F04103$-$2838 are strikingly
  similar to those of the local luminous infrared galaxy NGC~6240.
  Given the similarities in X-ray properties but differences in merger
  state and in infrared color and luminosity, objects like NGC~6240
  could conceivably be the precursors of ULIRGs like F04103$-$2838.

\end{enumerate}

\acknowledgements

We are grateful for the referee's comments and suggestions which helped
improve this paper.  Thanks are due to Drs. Chris Reynolds, Cole Miller, and Yuxuan Yang for useful
discussions.  This research is based on observations obtained with
{\it XMM-Newton}, an ESA science mission with instruments and
contributions directly funded by ESA Member States and NASA.  We made
use of the NASA/IPAC Extragalactic Database (NED), which is operated
by the Jet Propulsion Laboratory, Caltech, under contract with NASA.
We acknowledge support from NASA/{\it XMM-Newton} Guest Observer Program
under grant NNX06AF51G.

\clearpage



\begin{deluxetable}{cccccccccc}
\tabletypesize{\scriptsize}
\rotate 
\tablecolumns{10} \tablewidth{0pc}
  \tablecaption{Best-fit Parameters to Models A, B, \&
    C\tablenotemark{\dag}}
  \tablehead{\colhead{Model}&\colhead{A}&\colhead{A}&\colhead{A}&\colhead{A}&\colhead{B}&\colhead{B}&\colhead{B}&\colhead{B}&\colhead{C}\\
    \colhead{Parameters}&\colhead{15 cts bin$^{-1}$}&\colhead{5 cts
      bin$^{-1}$}&\colhead{3 cts bin$^{-1}$}&\colhead{unbinned}&\colhead{15 cts
      bin$^{-1}$}&\colhead{5 cts bin$^{-1}$}&\colhead{3 cts
      bin$^{-1}$}&\colhead{unbinned}&\colhead{unbinned}}
\startdata
  N$_{\rm H}\tablenotemark{a}$&0.20$^{+0.28}_{-0.16}$&0.00$^{+0.04}_{-0.00}$&0.00$^{+0.04}_{-0.00}$&0.00$^{+0.04}_{-0.00}$&0.30$^{+0.36}_{-0.22}$&0.00$^{+0.06}_{-0.00}$&0.00$^{+0.06}_{-0.00}$&0.00$^{+0.06}_{-0.00}$&0.19$^{+0.33}_{-0.19}$\\
  $\Gamma$&1.42$^{+0.61}_{-0.41}$&1.01$^{+0.21}_{-0.20}$&1.00$^{+0.21}_{-0.20}$&1.00$^{+0.21}_{-0.20}$&1.80$^{+0.90}_{-0.60}$&1.12$^{+0.25}_{-0.22}$&1.11$^{+0.26}_{-0.22}$&1.09$^{+0.27}_{-0.22}$&1.36$^{+0.97}_{-0.44}$\\
  E$_{\rm line}$\tablenotemark{b}& --- & --- & ---&---&6.57$^{+1.46}_{-1.20}$&6.37$^{+0.18}_{-0.17}$&6.42$^{+0.26}_{-0.29}$&6.43$^{+0.27}_{-0.28}$&6.43$^{+0.26}_{-0.26}$\\
  $\sigma$\tablenotemark{b}& --- & --- & --- &---&0.00$^{+1.20}_{-0.00}$&0.14$^{+0.20}_{-0.14}$&0.23$^{+0.28}_{-0.23}$&0.25$^{+0.25}_{-0.25}$&0.26$^{+0.24}_{-0.26}$\\
  EW\tablenotemark{b}& --- & --- & ---&---
  &1.96$^{+19.54}_{-1.95}$&1.33$^{+1.24}_{-0.89}$&1.37$^{+1.38}_{-1.01}$&1.39$^{+1.37}_{-1.02}$&1.62$^{+1.58}_{-1.12}$\\
kT\tablenotemark{b}& --- & --- & --- &---& --- & --- & --- &---&0.10$^{+0.03}_{-0.08}$\\
  Stat./$dof$\tablenotemark{c}&12.4/11&66.5/40&83.0/67&858.2/1958&8.6/8&58.8/37&77.1/64&852.1/1955&848.1/1953\\
  F$_{0.2-2 {\rm keV}}$
  (total)\tablenotemark{d}&0.96$^{+0.42}_{-0.71}$&1.07$^{+0.29}_{-0.21}$&1.07$^{+0.29}_{-0.21}$&1.05$^{+0.30}_{-0.20}$&2.93$^{+0.68}_{-1.40}$&1.08$^{+0.17}_{-0.22}$&1.08$^{+0.21}_{-0.19}$&1.06$^{+0.19}_{-0.21}$&2.17$^{+0.87}_{-0.87}$\\
  \indent \hspace{1.2cm} (AGN)\tablenotemark{d}& --- & --- & --- &---& 2.93$^{+0.68}_{-1.40}$&1.08$^{+0.17}_{-0.22}$&1.08$^{+0.21}_{-0.19}$ &1.06$^{+0.19}_{-0.21}$&1.64$^{+0.86}_{-0.50}$\\
  F$_{2-10 {\rm keV}}$
  (total)\tablenotemark{d}&3.96$^{+1.37}_{-2.57}$&4.70$^{+3.58}_{-2.04}$&4.73$^{+3.62}_{-2.12}$&4.85$^{+3.75}_{-2.10}$&3.65$^{+1.64}_{-1.02}$&4.57$^{+2.27}_{-3.05}$&4.60$^{+1.54}_{-3.27}$&4.76$^{+3.35}_{-2.75}$&4.56$^{+2.04}_{-0.98}$\\
  \indent \hspace{1.2cm} (AGN)\tablenotemark{d}& --- & --- & --- &---&3.65$^{+1.64}_{-1.02}$&4.57$^{+2.27}_{-3.05}$&4.60$^{+1.54}_{-3.27}$ &4.76$^{+3.35}_{-2.75}$&4.56$^{+1.92}_{-1.01}$\\
  L$_{0.2-2 {\rm keV}}$
  (total)\tablenotemark{e}&0.29$^{+0.12}_{-0.21}$&0.31$^{+0.09}_{-0.06}$&0.31$^{+0.09}_{-0.06}$&0.31$^{+0.09}_{-0.06}$&0.86$^{+0.20}_{-0.41}$&0.31$^{+0.05}_{-0.06}$&0.31$^{+0.06}_{-0.05}$&0.31$^{+0.05}_{-0.06}$&0.64$^{+0.26}_{-0.26}$\\
  \indent \hspace{1.2cm} (AGN)\tablenotemark{e}& --- & --- & --- &---&0.86$^{+0.20}_{-0.41}$&0.31$^{+0.05}_{-0.06}$&0.31$^{+0.06}_{-0.05}$ &0.31$^{+0.05}_{-0.06}$&0.48$^{+0.26}_{-0.15}$\\
  L$_{2-10 {\rm keV}}$
  (total)\tablenotemark{e}&1.16$^{+0.40}_{-0.76}$&1.39$^{+1.06}_{-0.60}$&1.40$^{+1.06}_{-0.62}$&1.44$^{+1.11}_{-0.62}$&1.07$^{+0.49}_{-0.30}$&1.35$^{+0.67}_{-0.90}$&1.36$^{+0.45}_{-0.96}$&1.40$^{+0.99}_{-0.81}$&1.35$^{+0.60}_{-0.29}$\\
  \indent \hspace{1.2cm} (AGN)\tablenotemark{e}& --- & --- & --- &---&1.07$^{+0.49}_{-0.30}$&1.35$^{+0.67}_{-0.90}$&1.36$^{+0.45}_{-0.96}$ &1.40$^{+0.99}_{-0.81}$&1.35$^{+0.57}_{-0.30}$\\
  \enddata

  \footnotesize 

  \tablenotetext{\dag} {Model A: Absorption$_{\rm Galactic} \times
    {\rm Absorption_{source}} \times {\rm PL}$. Model B:
    Absorption$_{\rm Galactic} \times {\rm Absorption_{source}} \times
    ({\rm PL + Line})$.  Model C: Absorption$_{\rm Galactic} \times [{\rm
    MEKAL} + {\rm Absorption_{source}} \times ({\rm PL +
    Line})]$, where MEKAL is the Mewe, Kaastra, \& Liedahl thermal
  plasma model (see the XSPEC manual for details), PL is a power-law
  model representing the AGN, Line is the Fe~K
    emission line with a Gaussian profile, Absorption$_{\rm Galactic}$
    is the absorption from N$_{\rm H, Galactic} = 2.45 \times
    10^{20}$~atoms~cm$^{-2}$, and Absorption$_{\rm source}$ is the
    intrinsic absorption within the source. }

  \tablenotetext{a}{Intrinsic (i.e. within the galaxy) column density in units of
    10$^{22}$~atoms~cm$^{-2}$.}

  \tablenotetext{b}{Fe~K line energy (rest frame), width, equivalent width, and
    thermal gas temperature all in keV.}

  \tablenotetext{c}{Fitting statistics per degrees of freedom. Cash
    statistics are used for unbinned spectra and spectra binned to at least 3 and 5 counts
    bin$^{-1}$, while $\chi^2$ statistics are used for spectra binned
    to at least 15 counts bin$^{-1}$.}

  \tablenotetext{d}{Absorption-corrected flux in units of 10$^{-14}$
    ergs~cm$^{-2}$~s$^{-1}$.  The AGN value includes the flux from both the power-law component and the iron line.}

  \tablenotetext{e}{Absorption-corrected luminosity in units of
    10$^{42}$ ergs~s$^{-1}$.  The AGN value includes the flux from both the power-law component and the iron line.}

\label{tab:binfits}
\end{deluxetable}


\clearpage

\begin{thebibliography}{}



\bibitem[Boller et al.(2003)]{boller03} Boller, Th., Keil, R.,
Hasinger, G., Costantini, E., Fujimoto, R., Anabuki, N., Lehmann, I.,
\& Gallo, L. 2003, A\&A, 411, 63

\bibitem[Braito et al.(2003)]{braito03} Braito, V., Franceschini, A.,
Della Ceca, R., Severgnini, P., Bassani, L., Cappi, M., Malaguti, G.,
Palumbo, G.G.C., Persic, M., Risaliti, G., \& Salvati, M. 2003, A\&A,
398, 107

\bibitem[Braito et al.(2004)]{braito04} Braito, V., Della Ceca, R.,
Piconcelli, E., Severgnini, P., Bassani, L., Cappi, M., Franceschini,
A., Iwasawa, K., Malaguti, G., Marziani, P., Palumbo, G.G.C., Persic,
M., Risaliti, G., \& Salvati, M. 2004, \aa, 420, 79

\bibitem[Bryant \& Scoville(1999)]{BS99} Bryant, P. M., \& Scoville,
  N. Z. 1999, \aj, 117, 2632

\bibitem[Cecil et al.(2002)]{cecil02} Cecil, G., Bland-Hawthorn, J.,
  \& Veilleux, S. 2002, \apj, 576, 745

\bibitem[Chen \& Zhang(2006)]{chen} Chen, P.S., \& Zhang, P. \aj, 131,
1942

\bibitem[Dasyra et al.(2006a)]{dasyra} Dasyra, K.M. et al. 2006a, ApJ,
  638 ,745

\bibitem[Dasyra et al.(2006b)]{dasyrab} Dasyra, K.M. et al. 2006b, ApJ,
  651, 835

\bibitem[Dasyra et al.(2007)]{dasyra07} Dasyra, K.M. et al. 2007, ApJ,
  657, 102

\bibitem[Dickey \& Lockman(1990)]{nh} Dickey \& Lockman, 1990, ARAA, 28, 215


\bibitem[Elvis et al.(1994)]{elvis} Elvis, M., Wilkes, B.J., McDowell,
J.C., Green, R.F., Bechtold, J., Willner, S.P., Oey, M.S., Polomski,
E., \& Cutri, R. 1994, ApJS, 95, 1.

\bibitem[Franceschini et al.(2003)]{frances} Franceschini, A., Braito,
V., Persic, M., Della Ceca, R., Bassani, L., Cappi, M., Malaguti, P.,
Palumbo, G.G.C., Risaliti, G., Salvati, M., \& Severgnini, P. 2003,
MNRAS, 343, 1181

\bibitem[Gonz\'{a}lez-Mart\'{i}n et al.(2006)]{gonzalez}
  Gonz\'{a}lez-Mart\'{i}n, O., Masegosa, J., M\'{a}rquez, I.,
  Guerrero, M.A., \& Dultzin-Hacyan, D. 2006, A\&A, 460, 45

\bibitem[Gehrels(1986)]{stat} Gehrels, N. 1986, \apj, 303, 336

\bibitem[Genzel et al.(1998)]{genzel} Genzel, R., et al. 1998, \apj,
498, 579


\bibitem[Ghisellini et al.(1994)]{ghm} Ghisellini, G., Harrdt, F., \&
Matt, G. 1994, \mnras, 267, 743


\bibitem[Grimes et al.(2005)]{grimes} Grimes, J.P., Heckman, T.,
Strickland, D., \& Ptak, A., 2005, \apj, 628, 187

\bibitem[Hopkins et al.(2005)]{hopkins} Hopkins, P.F., Hernquist, L.,
  Cox, T.J., Di Matteo, T., Martini, P., Robertson, B., \& Springel,
  V. 2005, \apj, 630, 705

\bibitem[Ishida(2004)]{ishida04} Ishida, C. M. 2004, PhD Thesis,
  University of Hawaii


\bibitem[Imanishi et al.(2003)]{iman} Imanishi, M., Terashima, Y.,
Anabuki, N., \& Nakagawa, T. 2003, \apjl, 596, L167

\bibitem[Imanishi \& Terashima(2004)]{iman04} Imanishi, M. \&
      Terashima, Y. 2004, \apj, 127, 758

\bibitem[Iwasawa et al.(1997)]{n1068} Iwasawa, K., Fabian, A.C., \&
Matt, G. 1997, \mnras, 289, 443

\bibitem[Iwasawa et al.(2005)]{arp220} Iwasawa, K., Sanders, D.B.,
Evans, A.S., Trentham, N., Miniutti, G., \& Spoon, H.W.W. 2005,
\mnras, 357, 565

\bibitem[Kim \& Sanders(1998)]{kim98} Kim, D.-C., \& Sanders,
D.B. 1998, \apjs, 119, 41

\bibitem[Kim et al.(1998)]{kim98b} Kim, D.-C., Veilleux, S., \&
Sanders, D.B. 1998, \apj, 508, 627

\bibitem[Kim et al.(2002)]{kim02} Kim, D.-C., Veilleux, S., \&
Sanders, D.B. 2002, \apjs, 143, 277

\bibitem[Kirsch et al.(2006)]{cal} Kirsch, M., \& EPIC Consortium,
2006, XMM-EPIC Status of Calibration and Data Analysis
[http://xmm.vilspa.esa.es/docs/documents/CAL--TN--0018.pdf]

\bibitem[Komossa et al.(2003)]{komossa} Komossa, S., Burwitz, V., Hasinger, G., Predehl, P., Kaastra, J.S., \& Ikebe, Y. 2003, \apjl, 582, L15

\bibitem[Krolik et al.(1994)]{krolik} Krolik, J.H., Madau, P., \&
Zycki, P.T. 1994, \apjl, 420, L57

\bibitem[Levenson et al.(2002)]{levenson} Levenson, N.A., Krolik, J.H., Zycki, P.T., Heckman, T.M., Weaver, K.A., Awaki, H., \& Terashima, Y. 2002, \apjl, 573, L81

\bibitem[Lutz et al.(1999)]{lutz99} Lutz, D., Veilleux, S., \& Genzel,
R. 1999, \apj, 517, L13

\bibitem[Maloney \& Reynolds(2000)]{malreynolds} Maloney, P.R. \&
Reynolds, C.S. 2000, \apjl, 545, L23

\bibitem[Netzer et al.(2005)]{netzer} Netzer, H., Lemze, D., Kaspi,
S., George, I.M., Turner, T.J., Lutz, D., Boller, Th., \& Chelouche,
D. 2005, \apj, 629, 739



\bibitem[Piconcelli et al.(2005)]{piconcelli} Piconcelli, E., Jimenez-Bailon, E., Guainazzi, M., Schartel, N., Rodriguez-Pascual, P. M., \& Santos-Lleo, M. 2005, A\&A, 432, 15

\bibitem[Ptak et al.(2003)]{ptak} Ptak, A., Heckman, T., Levenson,
N.A., Weaver, K., \& Strickland, D. 2003, \apj, 592, 787



\bibitem[Rupke et al.(2002)]{rupke02} Rupke, D.S., Veilleux, S., \&
Sanders, D.B. 2002, \apj, 570, 588

\bibitem[Rupke et al.(2005a)]{rupke05a} Rupke, D.S., Veilleux, S., \&
Sanders, D.B. 2005a, \apjs, 160, 87

\bibitem[Rupke et al.(2005b)]{rupke05b} Rupke, D.S., Veilleux, S., \&
Sanders, D.B. 2005b, \apjs, 160, 115

\bibitem[Rupke et al.(2005c)]{rupke05c} Rupke, D.S., Veilleux, S., \&
Sanders, D.B. 2005c, \apj, 632, 751

\bibitem[Rybicki \& Lightman(1979)]{rybicki} Rybicki, G.B., \&
Lightman, A.P., 1979, Radiative Processes in Astrophysics (New York
Wiley Interscience)

\bibitem[Sanders et al.(1989)]{san89} Sanders, D.B., Phinney, E.S., Neugebauer, G., Soifer, B.T., \& Matthews, K. 1989, \apj, 347, 29



\bibitem[Strickland \& Stevens(2000)]{winds} Strickland, D.K. \&
Stevens, I.R. 2000, MNRAS, 314, 511

\bibitem[Strickland et al.(2004a)]{s04a} Strickland, D.K., Heckman,
  T. M., Colbert, E. J. M., Hoopes, C. G., \& Weaver, K. A. 2004,
  \apj, 606, 829

\bibitem[Strickland et al.(2004b)]{s04b} Strickland, D.K., Heckman,
  T. M., Colbert, E. J. M., Hoopes, C. G., \& Weaver, K. A. 2004,
  \apjs, 151, 193

\bibitem[Surace \& Sanders(1999)]{surace} Surace, J.A., \& Sanders,
D.B. 1999, \apj, 512, 162

\bibitem[Taniguchi et al.(1999)]{taniguchi} Taniguchi, Y., Yoshino,
A., Ohyama, Y., \& Nishiura, S. 1999, \apj, 514, 660

\bibitem[Teng et al.(2005)]{teng} Teng, S.H., Wilson, A.S., Veilleux,
S., Young, A.J., Sanders, D.B., \& Nagar, N.M. 2005, \apj, 633, 664
(Paper I)


\bibitem[Terashima et al.(2002)]{tera} Terashima, Y., Iyomoto, N., Ho,
L.C., \& Ptak, A.F. 2002, \apjs, 139, 1

\bibitem[Veilleux et al.(1995)]{vei95} Veilleux, S., Kim, D.-C.,
Sanders, D.B., Mazzarella, J.M. \& Soifer, B.T., 1995, \apjs, 98, 171

\bibitem[Veilleux et al.(1997)]{vei97} Veilleux, S., Sanders, D.B., \&
Kim, D.-C. 1997, \apj, 484, 92

\bibitem[Veilleux et al.(1999a)]{vei99a} Veilleux, S., Kim, D.-C., \&
Sanders, D.B. 1999a, \apj, 522, 113

\bibitem[Veilleux et al.(1999b)]{vei99b} Veilleux, S., Sanders, D.B.,
\& Kim, D.-C. 1999b, \apj, 522, 139

\bibitem[Veilleux et al.(2002)]{vei02} Veilleux, S., Kim, D.-C., \&
Sanders, D.B. 2002, \apjs, 143, 315

\bibitem[Veilleux et al.(2006)]{vei06} Veilleux, S. et al., 2006,
  \apj, 643, 707

\bibitem[Vignati et al.(1999)]{vignati} Vignati, P. et al., 1999, A\&A,
  349, L57

\end{thebibliography}
\end{document}